# On Optimizing Rate Splitting in Laser-based Optical Wireless Networks

Khulood Alazwary, Ahmad Adnan Qidan, Taisir El-Gorashi and Jaafar M. H. Elmirghani
School of Electronic and Electrical Engineering University of Leeds, Leeds, United Kingdom
{elkal, A.A.Qidan, t.e.h.elgorashi, J.M.H.Elmirghani}@leeds.ac.uk

*Abstract*—Optical wireless communication (OWC) is a promising technology that has the potential to provide Tb/s aggregate rates. In this paper, interference management is studied in a Laser-based optical wireless network where vertical-cavity surface-emitting (VCSEL) lasers are used for data transmission. In particular, rate splitting (RS) and hierarchical rate splitting (HRS) are proposed to align multi-user interference, while maximizing the multiplexing gain of the network. Basically, RS serves multiple users simultaneously by splitting a message of a user into common and private messages, each message with a certain level of power, while on the other side users decode their messages following a specific methodology. The performance of the conventional RS scheme is limited in high density wireless networks. Therefore, the HRS scheme is developed aiming to achieve high rates where users are divided into multiple groups, and a new message called outer common message is used for managing inter-group interference. We formulate an optimization problem that addresses power allocation among the messages of the HRS scheme to further enhance the performance of the network. The results show that the proposed approach provides high achievable rates compared with the conventional RS and HRS schemes in different scenarios.

**Keywords**—Optical wireless networks, lasers, interference management and optimization.

## I. INTRODUCTION

The next generation of wireless communication networks must support the rapid increase in the use of the Internet in addition to the emergence of Internet of Things (IoT) applications. In this context, optical wireless communication (OWC) has been considered as an effective potential technology due to the massive licence-free bandwidth available in the optical band. OWC capacity to transmit video, data and voice can reach up to 20 Gb/s and beyond per link in indoor environments [1]. In addition, OWC provides high security compared with the current radio frequency (RF) wireless networks. Despite its features, OWC suffers from limitations in the availability of the channel link and multi-path propagation. However, the link budget of OWC has been improved by adopting various techniques including beam power, beam angle, and beam delay adaptations [2]. Furthermore, the energy efficiency of such OWC systems has been studied in [3] using light emitting diodes (LEDs) as transmitters. Recently, infrared lasers, namely vertical-cavity surface-emitting (VCSEL) lasers, have been proposed as alternative transmitters to LEDs due to their high modulation speed, spectral purity and low cost [4].

On the user side, different receivers have been studied to provide a wide field of view with low power consumption such as angle diversity receivers (ADRs), which are composed of multiple photodiodes, each photodiode points to a different direction [5].

For interference management in optical wireless networks, some previous works consider orthogonal transmission schemes that assign exclusive time or frequency resources to multiple users such as time division multiple access (TDMA), orthogonal frequency division multiple access (OFDMA), optical code division multiple access (OCDMA) [6], wavelength division multiple access (WDMA) [3] and space division multiple access (SDMA). On the other hand non-orthogonal multiple access (NOMA) is proposed to enhance the spectrum efficiency where multiple users can be served simultaneously over the same resources in the power domain [7]. Despite the low complexity of the aforementioned schemes, high advanced transmission schemes are highly required to support the unprecedented rates and user densities in OWC. In this sense, a multiple access scheme referred to as rate splitting (RS) is proposed in multiple-input single-output broadcast channel (MISO BC) RF systems to serve multiple users with the need for channel state information (CSI) at transmitters [8]. It is shown that the RS provides high spectral and energy efficiency compared with the conventional zero forcing (ZF) scheme. Additionally, RS is considered as an efficient scheme even under the imperfection of CSI. However, in the case of high density wireless networks, the performance of RS suffers degradation due to noise that results from interference cancelation. Given this point, a two-tier precoding RS scheme called hierarchical rate splitting (HRS) is particularly proposed to enhance the sum rate and alleviate CSI requirements [9]. In the HRS scheme, users are divided into multiple groups, and then, outer RS is applied to manage inter-group interference, while inner RS manages intra-group interference. In [4], the use of RS and HRS is proposed in optical wireless networks. It is shown that HRS provides higher data rate than RS.

In contrast to [4], in this paper, the impact of power allocation over outer common, inner common and private messages of the HRS scheme is investigated in a laser-based optical wireless network. We first define our system model, which is composed of multiple VCSELs serving multiple users. Then, the sum rate of the network is derived considering the implementation of the RS and HRS schemes. Finally, an optimization problem is formulated to allocate the power among the messages of the HRS such that the sum rate of the users is maximized. The results show that the proposed approach is more suitable for the laser-based optical wireless network considered than the conventional RS and HRS schemes, in which the power is allocated regardless of sum rate maximization.

The rest of this paper is organized as follows: Section II describes the system model, including the room, VCSEL and ADR configurations, while RS is discussed in Section III. In Section IV, the optimization problem is formulated. The simulation results are given and discussed in Section IV and the conclusions are stated in Section V.

## II. SYSTEM MODEL

We consider a downlink optical wireless network in an indoor environment that has dimensions (length × width × height) of 5 m × 5 m × 3 m, as shown in Fig. 1. It consists of $L$, $l = \{1,..,L\}$, optical APs distributed uniformly on the ceiling serving $K$, $k = \{1,..,K\}$, users. Each optical AP is composed of multiple VCSELs. The received power of the VCSEL is dictated by the beam waist $W_0$, the wavelength and the distance through which the beam travels. In the network considered, the transmitted signal can be written in vector form as

$$\mathbf{x} = \sum_{k=1}^{K} \sqrt{P_k} \mathbf{w}_k s_k, \quad (1)$$

where $\mathbf{s} = [s_1, s_2, ... s_K]$ is the data vector intended for $K$ users, and every $s_k$ has power allocation $\sqrt{P_k}$. Moreover, $w = [\mathbf{w}_1, \mathbf{w}_2 ... ..., \mathbf{w}_K]$ is the precoder vector. We assume that users are randomly distributed over the communication floor, which is a plane 0.85 m above the ground. Each user is equipped with ADR, which has multiple photodiodes that point to different directions, as shown in Fig. 2. All the photodiodes of ADR are connected to a single processing chain, consuming low power. It is worth mentioning that each photodiode has a narrow Field of View (FoV), and its direction is specified by its azimuth and elevation angles. Thus, each user has a wide FoV that guarantees its connection to most of the available APs. Given this point, the diffuse components of the optical channel can be neglected, for the sake of simplicity, as most of the received power is due to line-of-sight links. Accordingly, the signal received by a generic user $k$ can be expressed as

$$y_k = \sqrt{P_k}\mathbf{h}_k^H \mathbf{w}_k s_k + \underbrace{\sum_{j \neq k}^{K} \sqrt{P_j}\mathbf{h}_k^H \mathbf{w}_j s_j}_{multiuser\ interference} + z_k, \quad (2)$$

where $\mathbf{h}_k^H$ contains the channel responses between $L$ optical APs and user $k$. The second term of equation (2) is multi-user interference as multiple users are served simultaneously, and $z_k$ is defined as real valued additive white Gaussian noise with zero mean and variance $\sigma_k^2$ representing thermal noise and shot noise [10]. In this work, all the optical APs are connected to a central unit that has information on the distribution of users, and it controls the resources of the network. Moreover, CSI is available at the optical transmitters in order to serve the users considering the implementation of RS and HRS schemes.

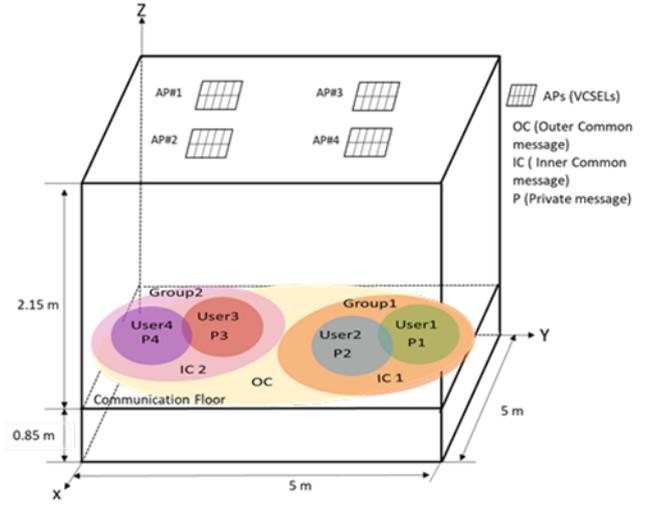

Fig 1. Room configuration and HRS illustrative example.

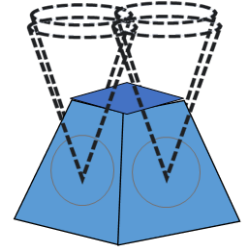

Fig 2. Angle Diversity Receiver.

### A. VCSEL Transmitter

VCSELs are a strong candidate in high-speed optical wireless communications due to their low cost and energy efficiency [11]. From the technical perspective, VCSELs utilize a multilayer semiconductor structure with distributed Bragg reflectors (DBR) having alternating high and low refractive index material such as GaAs and AlGaAs. The emission wavelengths of the GaAs/AlGaAs active medium VCSELs are in the range of 750 - 1550 nm. Furthermore, the wavelength of the VCSEL as well as its beam waist determinee the transmitted power. Note that, when using VCSELs as transmitters, the total power must be under eye safety constraints. In this work, we use multimode VCSELs, each mode is defined by two coefficients, $p \geq 0$ and $l \geq 0$, where $p$ and $l$ represent its radial and azimuthal order. The radial power profile of each mode is expressed as the superposition of Laguerre-Gaussian mode:

$$|u_{p,l}(r,z)|^2 = (A_p^l)^2 \cdot \frac{w_0^2}{w(z)^2} \cdot \left(\frac{2r^2}{w(z)^2}\right)^l \cdot \left(L_p^l\left(2\frac{r^2}{w(z)^2}\right)\right)^2 \cdot \exp\left(-\frac{2r^2}{w(z)^2}\right) \quad (3)$$

where $w_0$ indicates the beam waist at distance $z = 0$ from the aperture, and $w(z)$ represents the beam waist at a given distance $d$ which is given by

$$w(d) = w_0 \cdot \sqrt{1 + \left(\frac{\lambda d}{\pi n w_0^2}\right)^2} \quad (4)$$

The $L_p^l(x)$ indicates the generalised Laguerre polynomials which can be calculated as

$$L_p^l(x) = \sum_{m=0}^{p} (-1)^m \cdot \frac{(p+l)!}{(p-m)!(l+m)!m!} \cdot x^m \quad (5)$$

Moreover, $A_p^l$ is normalization constant for each mode which is used to ensure that the total mode power is 1, i.e.,

$$A_p^l = \frac{1}{w_0} \cdot \sqrt{\frac{2p!}{\pi(p+l)!}} \quad (6)$$

Therefore, the total output beam of the VCSEL is the combination of its modes that have optical power that depends on the structure and operating conditions of the VCSEL. Accordingly, the total radial mode power distribution can be determined as follows:

$$P(r,z) = \sum_{p=0}^{p_{max}} \sum_{l=0}^{l_{max}} a_{p,l} |u_{p,l}(r,z)|^2 \quad (7)$$

where $a_{p,l}$ is the coefficient of each mode power.

Table I. System Parameters

| Parameters | Configurations | | | |
|---|---|---|---|---|
| **VCSEL** | | | | |
| Beam waist, $W_0$ | 20 $\mu m$ | | | |
| Number of VCSELs per optical AP | 10 | | | |
| VCSEL wavelength | 850 $nm$ | | | |
| Number of optical transmitter units | 4 | | | |
| Optical transmitter locations (x, y, z) | (3.5m, 3.5m, 3m), (1.5m, 3.5m, 3m), (3.5 m, 1.5m, 3m), (1.5 m, 1.5 m, 3m) | | | |
| **Angle Diversity Receiver** | | | | |
| Responsivity | 0.4 A/W | | | |
| Number of photodetectors | 4 | | | |
| Area of the photodetector | 20 mm² | | | |
| Receiver noise current spectral density | 4.47 pA/√Hz | | | |
| Receiver bandwidth | 5 GHz | | | |
| Photodetector | *1* | *2* | *3* | *4* |
| Azimuth angles | 0° | 90° | 180° | 270 |
| Elevation angles | 60° | 60° | 60° | 60° |
| Field of view (FOV) | 25° | 25° | 25° | 25° |

## III. IMPLEMENTING RATE SPLITTING

In this section, we define the principles of RS and HRS schemes in a downlink MISO BC scenario considering our system model in section II.

### A. Rate Splitting

Basically, RS splits a message of a user into common and private messages. The common message is derived from a public codebook part where all users can decode it with minimum error. On the other hand, all the private messages of the users in the network are superimposed over the common message such that each user can decode its useful information following a specific interference cancelation process. In this context, a linear precoding technique can be used with CSI. That is, the transmitted signal to $K$ users can be expressed as:

$$\mathbf{x} = \sqrt{P_c}\mathbf{w}_c s_c + \sum_{k=1}^{K} \sqrt{P_k}\mathbf{w}_k s_k, \forall k \in K, \quad (8)$$

where $\mathbf{s} = [s_1, s_2, \dots s_K]$ is the data vector of private messages intended to $K$ users, while the $s_c$ is the common message transmitted to all the $K$ users. Moreover, $\mathbf{W} = [\mathbf{w}_1, \mathbf{w}_2 \dots \dots, \mathbf{w}_K]$ is designed using a zero forcing precoder, and $\mathbf{w}_c$ is the unit-norm precoding vector of the common message. Furthermore, a uniform power allocation approach can be performed for the sake of avoiding complexity where each private message is given a power level $P_k = P\alpha/K$, where $\alpha \in (0,1]$ is the fraction of the total power $P_t$ that is allocated to the private messages. Therefore, the remaining fraction of power allocated to the common message can be easily determined as $P_c = P(1-\alpha)$. At this point, the received signal of a generic user $k$ is given by

$$y_k = \sqrt{P_c}\mathbf{h}_k^H \mathbf{w}_c s_c + \sqrt{P_k}\mathbf{h}_k^H \mathbf{w}_k s_k + \sum_{j \neq k}^{K} \sqrt{P_j}\mathbf{h}_k^H \mathbf{w}_j s_j + z_k, \quad (9)$$

where $\mathbf{h}_k^H$ is the channel response vector between user $k$ and the optical transmitters and $z_k$ is defined in equation (2).

Each user follows two steps to decode its message: a user firstly decodes the common stream by treating all the private streams as noise. Ssecondly, the user decodes its private stream after subtracting the decoded common stream using Successive Interference Cancellation (SIC) from the original message received. Thus, the signal to interference and noise ratio (SINR) of the common and private messages denoted as $\gamma_k^c$ and $\gamma_k^p$ of user $k$ can be determined as

$$\gamma_k^c = \frac{P_c |\mathbf{h}_k^H \mathbf{w}_c|^2}{\sum_{j=1}^{K} P_j |\mathbf{h}_k^H \mathbf{w}_j|^2 + \sigma_k^2} \quad (10)$$

and

$$\gamma_k^p = \frac{P_k |\mathbf{h}_k^H \mathbf{w}_k|^2}{\sum_{j \neq k} P_j |\mathbf{h}_k^H \mathbf{w}_j|^2 + \sigma_k^2} \quad (11)$$

respectively. Accordingly, the achievable rate of the common message is given by

$$R_c^{RS} = \log_2(1 + \gamma^c), \quad (12)$$

where $\gamma^c = \min_k\{\gamma_k^c\}$ guarantees that the common message is successfully decoded by all the users. At user $k$, the achievable rate of the private message is $R_k^{RS} = \log_2(1+\gamma_k^p)$. Thus, the sum rate of the private messages of all the users can be obtained as

$$R_p^{RS} = \sum_{k=1}^{K} R_k^{RS} = \sum_{k=1}^{K} \log_2\left(1+\gamma_k^p\right). \quad (13)$$

Consequently, the sum rate of RS is given by

$$R_{\text{sum}}^{RS} = R_c^{RS} + R_p^{RS} \quad (14)$$

It is worth mentioning that the conventional RS scheme provides high data rates in small size networks, and its performance might suffer degradation as the number of users increases due to lack of accurate CSI at transmitters. Therefore, a new scheme is developed with the aim of providing high sum rates in such large size networks, with a two-tier precoder

*B. Hierarchical Rate Splitting (HRS)*

HRS is proposed to relax CSIT requirements, serving a large number of users at a higher data rate compared to the RS scheme. Specifically, users in the network are divided into multiple groups given by $G$, where each group contains a set of users that are located close to each other. Note that, the K-means clustering algorithm can be used to partition users into multiple sets based on the distance. In this context, HRS exploits the spatial correlation matrix known at the transmitter. The implementation of HRS scheme involves the use of two RS schemes referred to as outer and inner RS for managing inter-group interference and intra-group interference, respectively. The outer RS transmits an outer common message that can be decoded by all users aligning the inter-group interference. On the other hand, the inner RS transmits inner common messages given by the number of groups to align multi-user interference within each group where each inner message contains private messages intended to a certain set of users Therefore, the transmitted signal of HRS can be expressed as

$$\mathbf{x} = \sqrt{P_{oc}}\mathbf{w}_{oc}s_{oc} + \sum_{g=1}^{G} \mathbf{B}_g\left(\sqrt{P_{ic,g}}\mathbf{w}_{ic,g}s_{ic,g} + \sum_{k=1}^{K_g}\sqrt{P_{gk}}\mathbf{w}_{gk}s_{gk}\right) \quad (15)$$

where $\sum_{g=1}^{G} K_g = K$, $\mathbf{B}_g$ is the outer precoder and $\mathbf{W}_g = [\mathbf{w}_{g1}, \mathbf{w}_{g2}, \ldots, \mathbf{w}_{gK}]$ is the inner precoder. Moreover, $\mathbf{s}_g$ is a vector of the messages intended to the users belonging to group $g$. The power allocated to each message is jointly determined by $\beta$ and $\alpha$, where $\beta \in (0,1]$ is the fraction of the total power allocated to the messages of each group, while $\alpha \in (0,1]$ determines the fraction of power allocated to the private messages of each group. That is, the power of the outer common message is $P_{oc} = P(1-\beta)$, the power of the inner common message is $P_{ic,g} = \frac{P\beta}{G}(1-\alpha)$ and the power of the private message is $P_{gk} = \frac{P\beta}{K}\alpha$. Then, the received signal at user $k$ in group $g$ is given by

$$y_{gk} = \sqrt{P_{oc}}\mathbf{h}_{gk}^H\mathbf{w}_{oc}s_{oc} + \sqrt{P_{ic,g}}\mathbf{h}_{gk}^H\mathbf{B}_g\mathbf{w}_{ic,g}s_{ic,g}$$
$$+\sqrt{P_{gk}}\mathbf{h}_{gk}^H\mathbf{B}_g\mathbf{w}_{gk}s_{gk} + \sum_{j\neq k}^{K_g}\sqrt{P_{gj}}\mathbf{h}_{gk}^H\mathbf{B}_g\mathbf{w}_{gj}s_{gj} \quad (16)$$
$$+\sum_{l\neq g}^{G}\mathbf{h}_{gk}^H\mathbf{B}_l\mathbf{W}_l\mathbf{P}_l\mathbf{s}_l + z_{gk}$$

The decoding procedure of HRS starts by removing the outer and inner common messages from the received signal through performing SIC. Then, each user belonging to a group can decode its private message successfully. The Signal to Interference plus Noise Ratios of the outer and inner common messages and the private message of user $k$ can be expressed as

$$\gamma_{gk}^{oc} = \frac{P_{oc}\left|\mathbf{h}_{gk}^H\mathbf{w}_{oc}\right|^2}{\sum_{l=1}^{G}\sum_{j=1}^{K_g}P_{lj}\left|\mathbf{h}_{gk}^H\mathbf{B}_l\mathbf{w}_{lj}\right|^2 + \sum_{l=1}^{G}P_{ic,l}\left|\mathbf{h}_{gk}^H\mathbf{B}_l\mathbf{w}_{ic,l}\right|^2 + \sigma_{gk}^2} \quad (17)$$

$$\gamma_{gk}^{ic} = \frac{P_{ic,g}\left|\mathbf{h}_{gk}^H\mathbf{B}_g\mathbf{w}_{ic,g}\right|^2}{\sum_{l=1}^{G}\sum_{j=1}^{K_g}P_{lj}\left|\mathbf{h}_{gk}^H\mathbf{B}_l\mathbf{w}_{lj}\right|^2 + \sum_{l\neq g}^{G}P_{ic,l}\left|\mathbf{h}_{gk}^H\mathbf{B}_l\mathbf{w}_{ic,l}\right|^2 + \sigma_{gk}^2} \quad (18)$$

$$\gamma_{gk}^p = \frac{P_{gk}\left|\mathbf{h}_{gk}^H\mathbf{B}_g\mathbf{w}_{gk}\right|^2}{\sum_{l=1}^{G}\sum_{j=1}^{K_g}P_{lj}\left|\mathbf{h}_{gk}^H\mathbf{B}_l\mathbf{w}_{lj}\right|^2 - \Lambda + \sum_{l\neq g}^{G}P_{ic,l}\left|\mathbf{h}_{gk}^H\mathbf{B}_l\mathbf{w}_{ic,l}\right|^2 + \sigma_{gk}^2} \quad (19)$$

Respectively, where $\Lambda = P_{gk}\left|\mathbf{h}_{gk}^H\mathbf{B}_g\mathbf{w}_{gk}\right|^2$. Moreover, the achievable rate of the outer common message is given by

$$R_{oc}^{HRS} = \log_2(1+\gamma^{oc}), \gamma^{oc} = \min_{g,k}\{\gamma_{gk}^{oc}\}.$$

While, the sum rate of the inner common messages is given as $R_{ic}^{HRS} = \sum_{g=1}^{G}\log_2\left(1+\gamma_g^{ic}\right), \gamma_g^{ic} = \min_k\{\gamma_k^{ic}\}$, and the sum rate of the private messages is given by

$R_p^{HRS} = \sum_{g=1}^{G}\sum_{k=1}^{K_g}\log_2\left(1+\gamma_{gk}^p\right)$. Therefore, the sum rate of HRS is expressed as

$$R_{\text{sum}}^{HRS} = R_{oc}^{HRS} + R_{ic}^{HRS} + R_p^{HRS}. \quad (20)$$

For illustrative purposes, Fig.1. describes an example of a small optical wireless network, in which the transmission to two groups, each group with two users, is carried out using HRS. It can be seen that the outer RS transmits one outer common stream (OC), which is intended for both groups, and the inner RS transmits two inner common streams (IC 1) and (IC 2) for group 1 and group 2, respectively. Each inner common stream contains two private streams for two users, and therefore, four private messages are transmitted in total. It is worth pointing out that each user follows three steps to decode its information. Focussing on user 1, the outer common stream OC is decoded first treating the inner common streams IC and private streams as noise. After that, user 1 decodes its inner common stream by subtracting the decoded outer common stream from the original signal received using SIC and treating the private streams as noise. Finally, user 1 can decode its private stream after subtracting the decoded outer and inner common streams and managing other private messages intended for other users as noise.

IV. OPTIMIZATION PROBLEM

Power allocation is a curial issue in optical wireless communications. Therefore, an approach for managing the power of the network effectively is needed in such optical wireless networks. In this section, an optimization problem is formulated in the context of power allocation considering the implementation of the HRS scheme, which has the potential to provide sum rates higher than the conventional RS scheme. Recall that the sum rate of the HRS scheme is given by the sum of outer common, inner common and private rates. It is worth pointing out that the power levels allocated to these messages are related to each other. Therefore, an optimal power allocation approach is difficult to design due to high complexity. Therefore, we define a utility-based objective function that aims to maximize the overall sum rate of the network under power constraints, providing an effective solution. Note that, the users are divided into multiple groups, each with a unique set of users, based on the closest distance in order to apply the HRS scheme. In this sense, a high SNR case can be considered, i.e., low inter-group interference, for the sake of simplicity, where the outer common message is no longer needed, and the HRS scheme works as a RS scheme in each group. Given this point, the optimization problem can be formulated as follows:

$$\max_{p} \sum_{l \in L} \sum_{g \in G} \varphi \left( R_{ic,g}^{HRS}(P_{ic}) + \sum_{k \in K_g} R_{p,k}^{HRS}(P_k) \right)$$

$$\text{s.t. } P_{min} \leq \sum_{l \in L} \sum_{g \in G} \sum_{k \in K_g} P_k \leq P_{max}$$

$$\sum_{l \in L} \sum_{g=1}^{G} \left( P_{ic,g} + \sum_{k=1}^{K_g} P_k \right) \leq P_{t,b} \quad (21)$$

$$R_{sum}^{HRS} \geq R_{min}^{HRS}$$

$$k \in K, g \in G,$$

where $\varphi(.)$ is a monotonically increasing, strictly concave and continuously differentiable function, which can be defined as a logarithmic function, i.e. $\varphi(.) = \log(.)$, to achieve proportional fairness among the groups of the HRS scheme. The fist constraint defines the power allocated to the private messages where $P_{min}$ and $P_{max}$ are the minimum power and the maximum power allocated for the private messages, respectively. Interestingly, the power allocated to the private messages goes towards the minimum value as intra- group interference increases due to the imperfection of the channel state information, otherwise the power of the private messages increases within the range specified. Moreover, after allocating the power for the private messages, the remaining power is allocated for the common messages that further enhance the sum rate of the HRS scheme. The second constraint in equation (21) guarantees that the total power consumed is equal to or less than the total power budget, which is limited due to eye safety constraints. The third constraint ensures that the sum rate of the network is higher than the minimum rate required by the network. The problem in (21) is defined as a non-convex optimization problem, which is complicated to solve. In this sense, the successive convex approximation algorithm can be applied to solve it at the cost of providing a sub-optimal solution. It is proven in many previous works that the successive convex approximation algorithm can provide a solution after a set of iterations that satisfies the Karush- Kuhn-Tucker (KKT) conditions of the original optimization problem [21].

## V. SIMULATION RESULTS

We consider an indoor environment with dimensions 5 m × 5 m × 3 m to evaluate the performance of the proposed approach compared with some benchmark schemes. On the ceiling, $L = 4$ optical APs are deployed, each with 10 VCSELs, to serve $K$ users distributed on the receiving plane randomly. Recall that, each user is equipped with ADR that provides a wide FoV ensuring full connectivity with most of the available optical APs. The rest of the parameters are mentioned in Table I.

In Fig. 3, the sum rate is shown against a range of SNR values from 5 dB to 35 dB for the proposed approach. The OMA, conventional RS and HRS schemes achieve 5 bits/s/Hz, 10 bits/s/Hz and 26 bits/s/Hz sum rates at 15 dB SNR, respectively. On the other hand, the proposed approach achieves 39 bits/s/Hz sum rate at the same value of the SNR as the power is allocated among users with the aim of maximizing the sum rate of the network (see the optimization problem in (21)). Therefore, the proposed approach is considered an effective solution even if the users experience a low SNR, which can be caused by the imperfection of the CSI at transmitters. Note that, the sum rates of all the schemes increase with the SNR where each user can successfully eliminate the interference at high SNR values, achieving a high user rate.

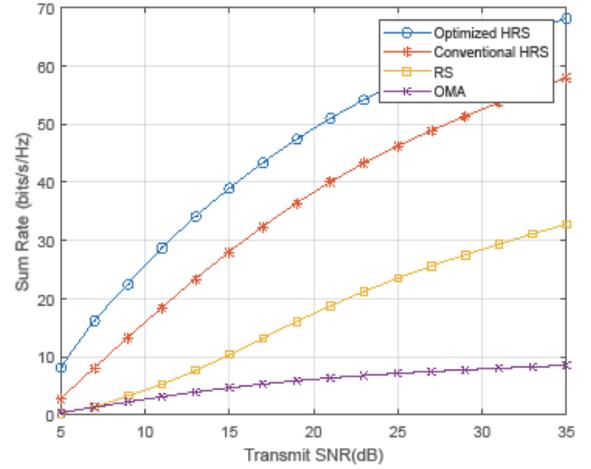

Figure 3. Sum rate versus SNR

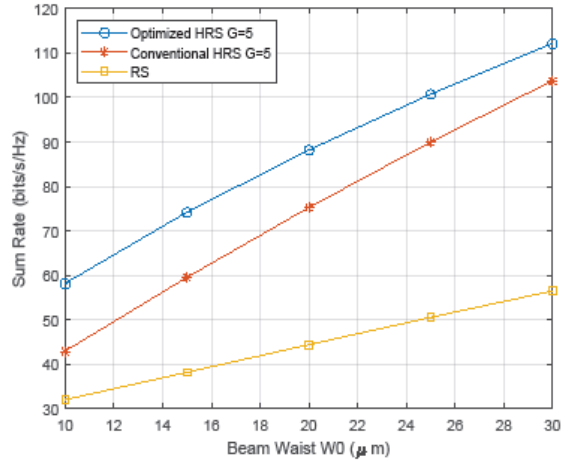

Figure 4. Sum rate versus beam waist $W_0$

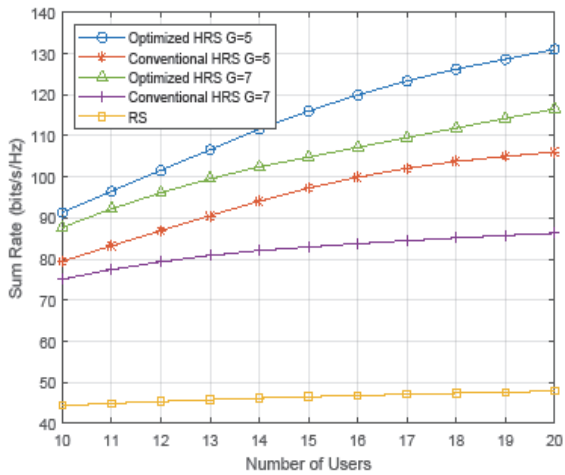

Figure 5. Sum Rate versus Number of Users

In Fig. 4, the sum rates of the proposed approach and the conventional RS and HRS schemes are depicted versus different values of the VCSEL beam waist $W_0$. It can be seen that performing the optimization problem in (21) results in enhancing the sum rate of the users considerably compared with the performance of the RS and HRS schemes where the power is simply allocated in uniform fashion among the messages of the users. The figure also shows that the sum rate of the users increases with the beam waist $W_0$ reagrdless of the scheme considered, due to the fact that the increase of the beam waist results in a high received power and less beam divergence. However, eye safety must be taken into account.

In Fig.5, the sum rate of the proposed approach is depicted against the number of users. It is shown that the proposed approach achieves higher sum rates compared with the conventional RS and HRS schemes in all the scenarios considered. It is expected that the sum rates of all the schemes increase with the number of users due to the channel gain differences among those users. However, the lack of the resource among users must be taken into consideration. It is worth mentioning that the number of the assigned groups highly affects the sum rate of the proposed approach due to the trade-off among the number of groups, noise and the complexity within each group. As future work, an optimization problem can be formulated to jointly find the optimal number of groups and power allocation.

## VI. CONCLUSIONS

In this paper, the performance of the RS and HRS schemes is evaluated in a laser-based optical wireless network. First, the system model is defined considering an optical wireless network with multiple lasers serving multiple users. Then, two transmission schemes, RS and HRS, are derived to align multi-user-interference. In these schemes, the power is allocated among the messages of users uniformly to avoid complexity. However, an advanced power allocation approach is needed. Therefore, an optimization problem is formulated to allocate the power among users, while the maximization of the sum rate of the network is taken into consideration. The results show that the proposed approach is more suitable for the optical wireless network than the conventional RS and HRS schemes where each user can achieve a high data rate even in a low SNR scenario.

ACKNOWLEDGMENTS

The authors would like to acknowledge funding from the Engineering and Physical Sciences Research Council (EPSRC) INTERNET (EP/H040536/1), STAR (EP/K016873/1) and TOWS (EP/S016570/1) projects. KDA would like to thank King Abdulaziz University in the Kingdom of Saudi Arabia for funding her PhD scholarship. All data are provided in full in the results section of this paper.

REFERENCES

[1] F. E. Alsaadi, M. A. Alhartomi, and J. M. H. Elmirghani, "Fast and efficient adaptation algorithms for multi-gigabit wireless infrared systems," J. Light. Technol., vol. 31, no. 23, pp. 3735–3751, 2013, doi: 10.1109/JLT.2013.2286743.

[2] M. T. Alresheedi and J. M. H. Elmirghani, "10 Gb/s indoor optical wireless systems employing beam delay, power, and angle adaptation methods with imaging detection," J. Light. Technol., vol. 30, no. 12, pp. 1843–1856, 2012, doi: 10.1109/JLT.2012.2190970.

[3] O. Z. Alsulami et al., "Optimum resource allocation in optical wireless systems with energy-efficient fog and cloud architectures," Philos. Trans. R. Soc. A Math. Phys. Eng. Sci., vol. 378, no. 2169, Apr. 2020, doi: 10.1098/rsta.2019.0188.

[4] K. Alazwary, A. A. Qidan, T. El-Gorashi, and J. M. H. Elmirghani, "Rate splitting in VCSEL-based optical wireless networks," 2021 6th Int. Conf. Smart Sustain. Technol. Split. 2021, pp. 1–5, 2021, doi: 10.23919/SpliTech52315.2021.9566354.

[5] A. Al-Ghamdi and J. M. H. Elmirghani, "Optimization of a triangular PFDR antenna in a fully diffuse OW system influenced by background noise and multipath propagation," IEEE Trans. Commun., vol. 51, no. 12, pp. 2103–2114, 2003, doi: 10.1109/TCOMM.2003.820758.

[6] J. M. H. Elmirghani and R. A. Cryan, "New PPM-CDMA hybrid for indoor diffuse infrared channels," Electron. Lett., vol. 30, no. 20, pp. 1646–1647, 1994, doi: 10.1049/el:19941136.

[7] M. K. Aljohani et al., "NOMA visible light communication system with angle diversity receivers," in International Conference on Transparent Optical Networks, Jul. 2020, vol. 2020-July, doi: 10.1109/ICTON51198.2020.9203212.

[8] Y. Mao, B. Clerckx, and V. O. K. Li, "Energy Efficiency of Rate-Splitting Multiple Access, and Performance Benefits over SDMA and NOMA," Proc. Int. Symp. Wirel. Commun. Syst., vol. 2018-Augus, no. Ic, pp. 1–5, 2018, doi: 10.1109/ISWCS.2018.8491100.

[9] M. Dai, B. Clerckx, D. Gesbert, and G. Caire, "A Rate Splitting Strategy for Massive MIMO with Imperfect CSIT," IEEE Trans. Wirel. Commun., vol. 15, no. 7, pp. 4611–4624, Jul. 2016, doi: 10.1109/TWC.2016.2543212.

[10] J. M. Kahn and J. R. Barry, "Wireless infrared communications," Proc. IEEE, vol. 85, no. 2, pp. 265–298, 2002, doi: 10.1109/5.554222.

[11] C.-H. Yeh, Y.-C. Yang, C.-W. Chow, Y.-W. Chen, and T.-A. Hsu, "VCSEL and LED Based Visible Light Communication System by Applying Decode-and-Forward Relay Transmission," J. Light. Technol., vol. 38, no. 20, pp. 5728–5732, Jun. 2020, doi: 10.1109/jlt.2020.3003352.